\documentclass[final            
  ]
  {aipproc}

\layoutstyle{6x9}

\newcommand{\ve}[1]{\ensuremath{\mathbf{#1}}}

\newcommand{\n}[1]{\ensuremath{|\mathbf{#1}|}}

\newcommand{\nMF}{\ensuremath{n_t^\textrm{MF}}}

\newcommand{\PMF}{\ensuremath{P_t^\textrm{MF}}}

\begin{document}
\graphicspath{{plot/}}
\title{Spectral functions for medium-sized nuclei}


\author{Artur M. Ankowski}{
    address={Institute of Theoretical Physics\\ University of Wroc{\l}aw\\ pl. M. Borna 9, 50-204 Wroc\l aw, Poland}
    }

\author{Jan T. Sobczyk}{
  address={Institute of Theoretical Physics\\ University of Wroc{\l}aw\\ pl. M. Borna 9, 50-204 Wroc\l aw, Poland} 
  }

\begin{abstract}
The spectral functions for calcium and argon are constructed. It is verified that their predictions for the quasielastic electron-nucleus cross sections in the energy range $\sim$1~GeV agree with the data. The argon spectral function is then used to obtain the quasielastic neutrino-nucleus cross section.
\end{abstract}

\maketitle

The description of lepton-nucleus interactions in the $1$~GeV energy region is based on the impulse approximation (IA). The most reliable implementation of the IA is that of the spectral function (SF). Exact calculations of the SF exist only for the lightest nuclei, whereas well founded approximations for carbon and oxygen~\cite{ref:Benhar&Fabrocini&Fantoni&Sick,ref:Benhar&Farina&Nakamura}. In this note, we report results on modeling of the SFs for heavier nuclei like~$^{40}_{20}$Ca and~$^{40}_{18}$Ar.

Our primary motivation was to provide a good description of medium-sized nuclei by their SFs and, as a result, to improve accuracy of the predicted neutrino-nucleus cross sections with respect to the Fermi gas (FG) model for neutrino energy around $1$~GeV. We focus on quasielastic (QE) scattering and therefore verify obtained SFs in the region of energy transfer up to the QE peak.

\paragraph{\bf Model description}
Every SF is divided into two parts: the correlated and the mean field (MF) one. Analytical formula for the correlated part of the SF is given in Ref.~\cite{ref:Kulagin&Petti}. The MF part is modeled on the simplifying assumption that the level widths does not depend on momentum. The obtained expression contains a contribution from each shell-model state~$\alpha$ of energy $E_{\alpha}$ occupied by $N_t^{\alpha}$ nucleons:
\begin{equation}\label{eq:PMF}
\PMF(\ve p,E)=\nMF(\ve p)\frac1{N_t}\sum_{\alpha} N_t^{\alpha}F_\alpha(E_{\alpha}-E)~.%
\end{equation}
In the above, the MF part of the momentum distribution is denoted by~\nMF~and the number of nucleons of isospin~$t$ by $N_t$. As the function $F_\alpha$, describing width of level $\alpha$, we use the Gaussian distribution. For further details, see~\cite{ref:Ankowski&Sobczyk_GSF}.

In the numerical computations, we use the momentum distributions from Ref.~\cite{ref:Bisconti&Arias&Co} and recent parametrization of the electro{\-}magnetic form factors~\cite{ref:BBBA05}. To evaluate the off-shell hadronic current matrix elements, de Forest's prescription~\cite{ref:de Forest} is adopted. In the electromagnetic case, a procedure to restore the current conservation is applied. 

Two separate final state interactions (FSI) effects are included: Pauli blocking and reinteractions of the struck nucleon described by means of the optical potential~\cite{ref:FSI_Nakamura&Seki&Sakuda}.

\begin{figure}
    \begin{minipage}[l]{0.331\textwidth}
        \includegraphics[width=4.9cm]{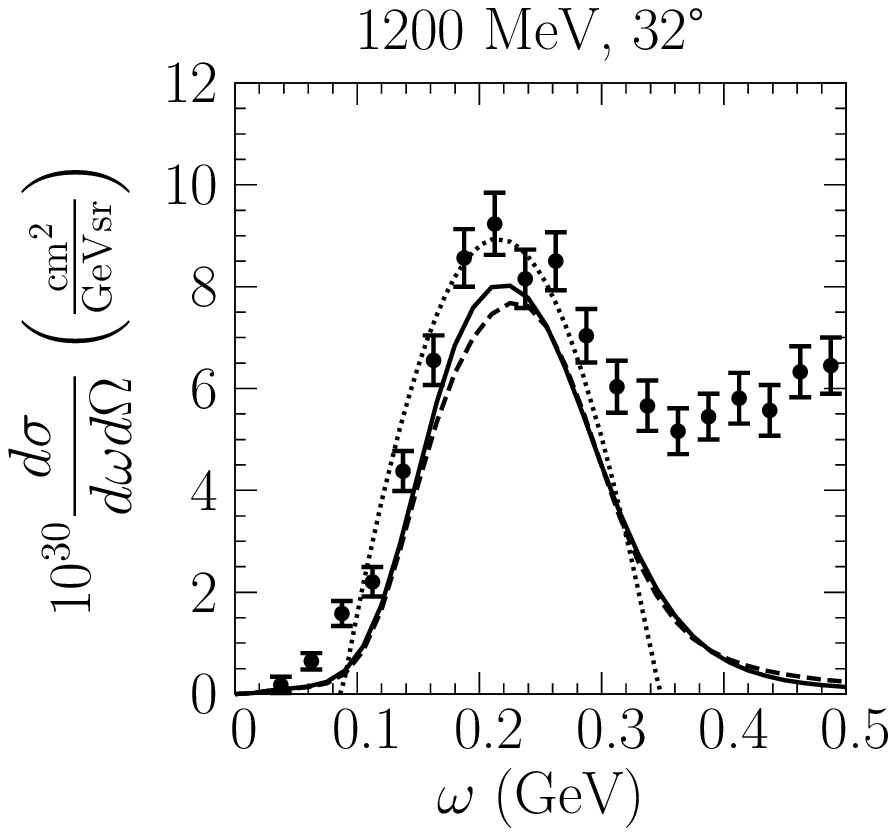}%
    \end{minipage}
    \begin{minipage}[c]{0.331\textwidth}
        \includegraphics[width=4.9cm]{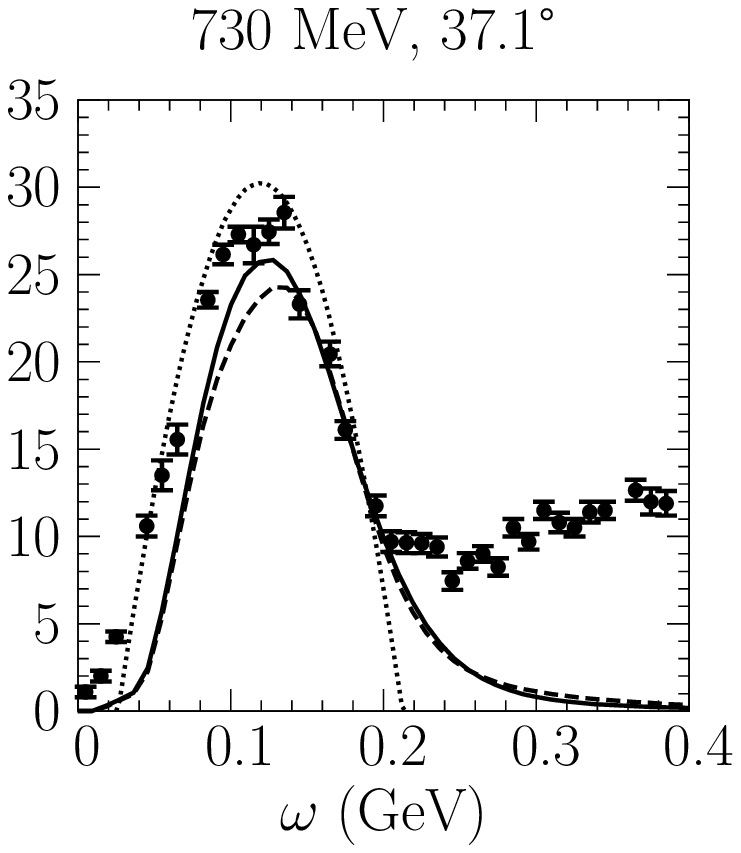}%
    \end{minipage}
    \begin{minipage}[r]{0.331\textwidth}
        \includegraphics[width=4.9cm]{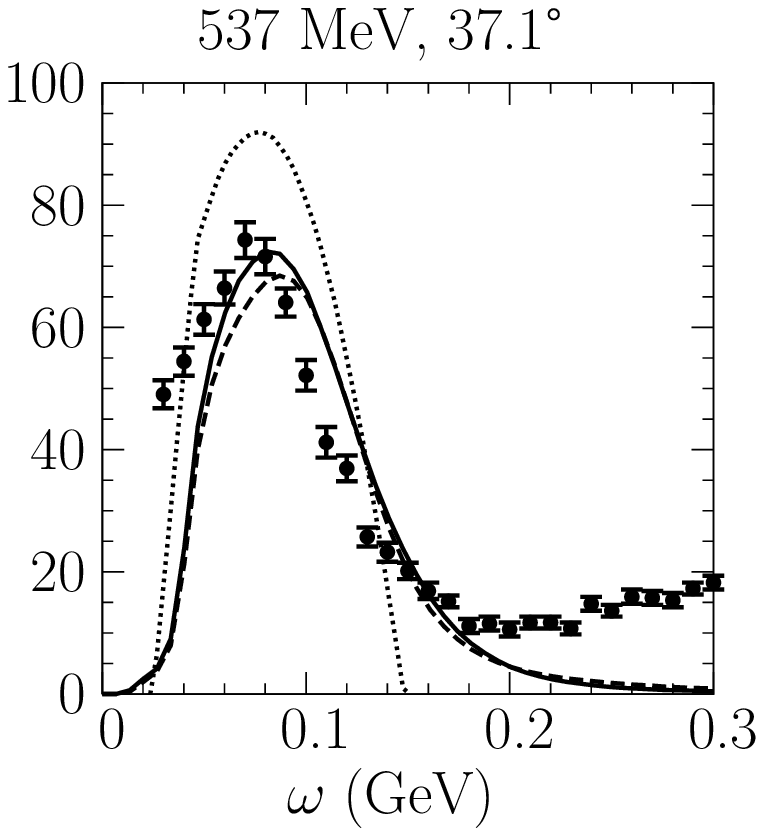}%
    \end{minipage}
\caption{\label{fig:Oxy}Differential cross section of the process~$^{16}$O$(e,e')$ for fixed scattering angle. Results for our model (solid line) are compared to the Benhar SF~\cite{ref:Benhar&Farina&Nakamura} with the same FSI (dashed line) and to the Fermi gas model (dotted line). The experimental data are from~\cite{ref:Anghinolfi,ref:O'Connell}. The values of momentum transfer at the peaks are 638~MeV (for beam energy 1200~MeV), 441~MeV (for 730~MeV), and 325~MeV (for 537~MeV).}
\end{figure}
\begin{figure}
    \begin{minipage}[l]{0.331\textwidth}
    \flushleft
        \includegraphics[width=4.9cm]{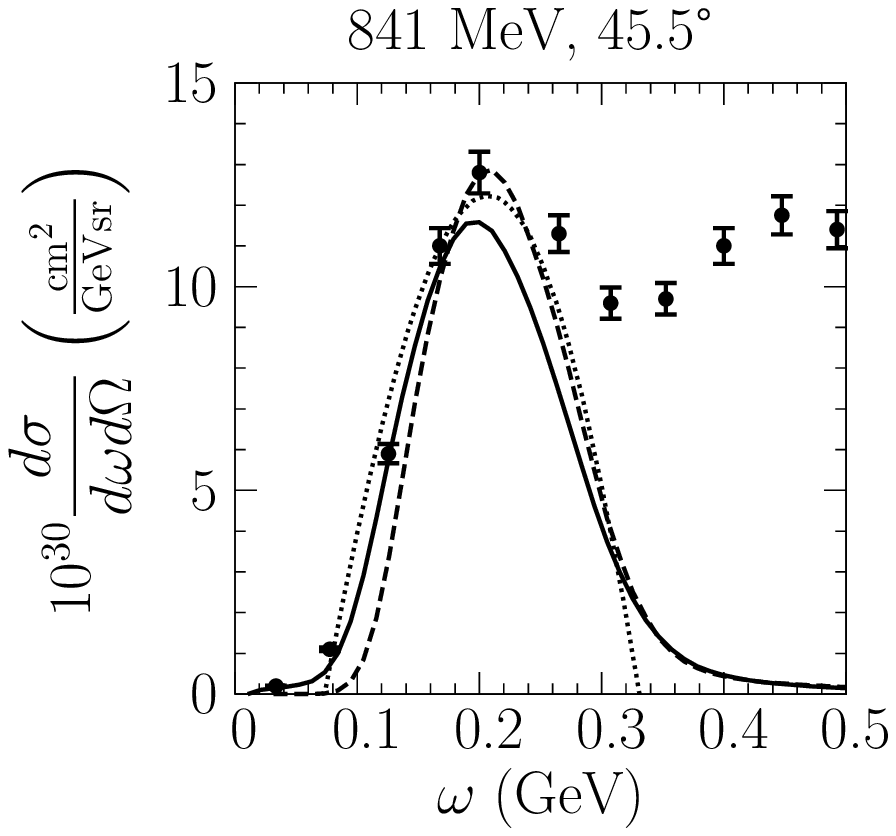}%
    \end{minipage}
    %
    \begin{minipage}[c]{0.331\textwidth}
        \includegraphics[width=4.9cm]{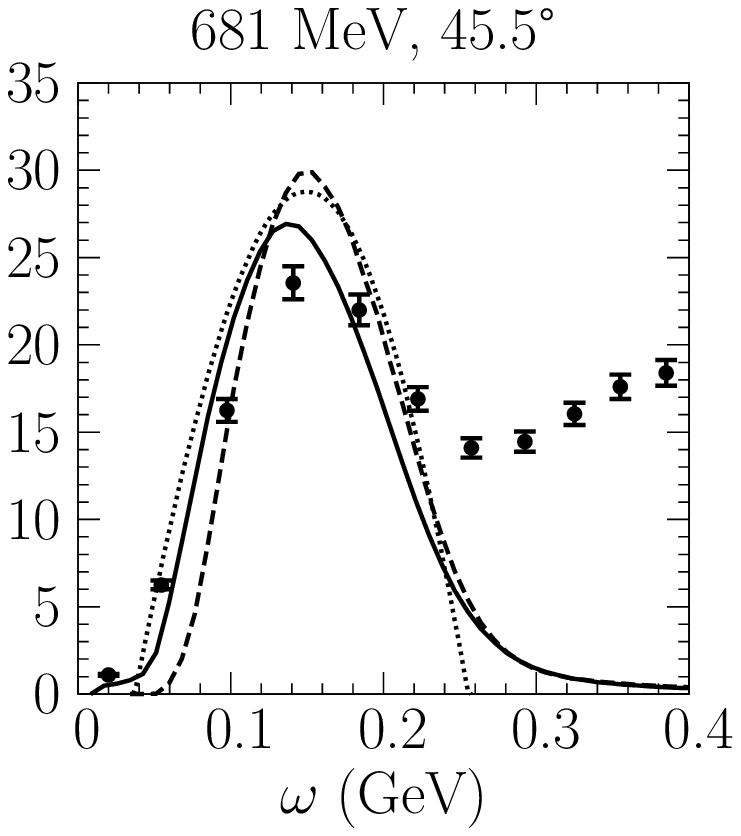}%
    \end{minipage}
    %
    \begin{minipage}[r]{0.331\textwidth}
        \includegraphics[width=4.9cm]{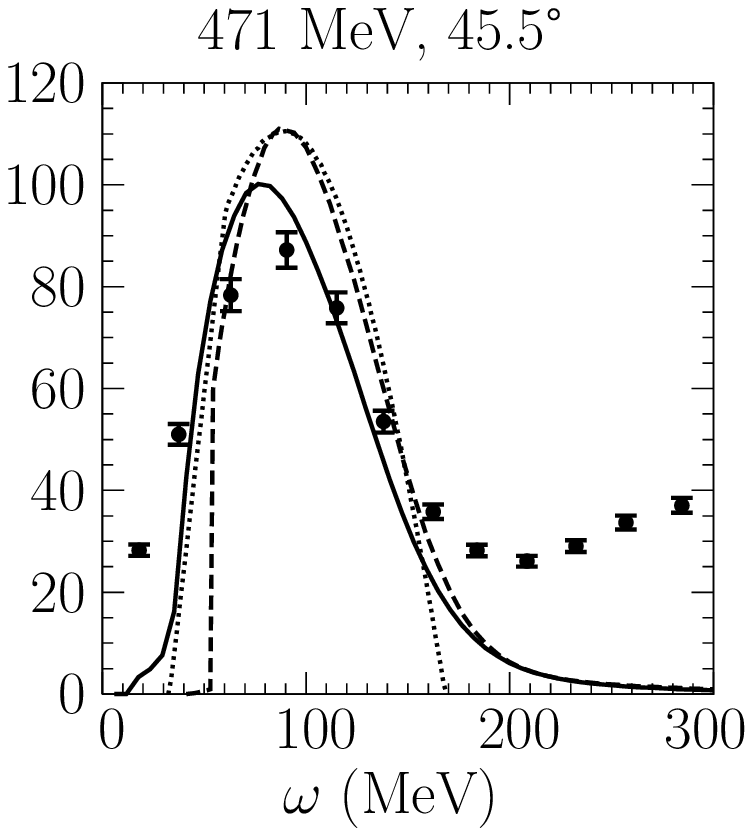}%
    \end{minipage}
\caption{\label{fig:Ca}Differential cross section of~$^{40}$Ca$(e,e')$ scattering at angle $45.5^\circ$. Calculations within our model (solid line) are compared to the results of Butkevich and Mikheyev~\cite{ref:Butkevich&Mikheyev} (dashed line), the Fermi gas model (dotted line), and the experimental data~\cite{ref:Williamson}. The corresponding values of~\n q~at the peaks are 602~MeV (for beam energy 841~MeV), 490~MeV (for 681~MeV), and 342~MeV (for 471~MeV).}
\end{figure}
\begin{figure}
    \begin{minipage}[l]{0.49\textwidth}
    \centering
        \includegraphics[width=6.1cm]{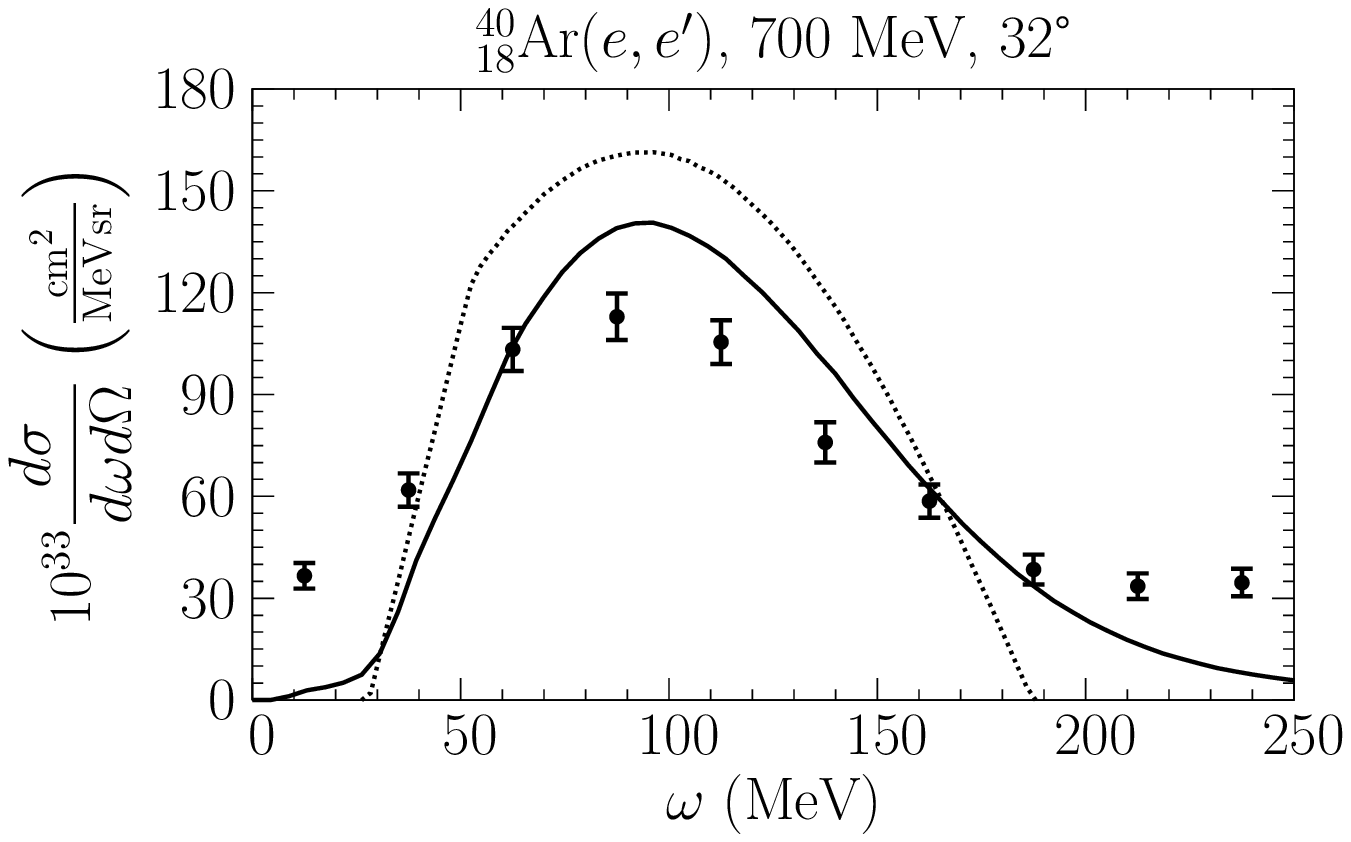}%
    \end{minipage}
    %
    \begin{minipage}[r]{0.49\textwidth}
    \centering
        \includegraphics[width=6.15cm]{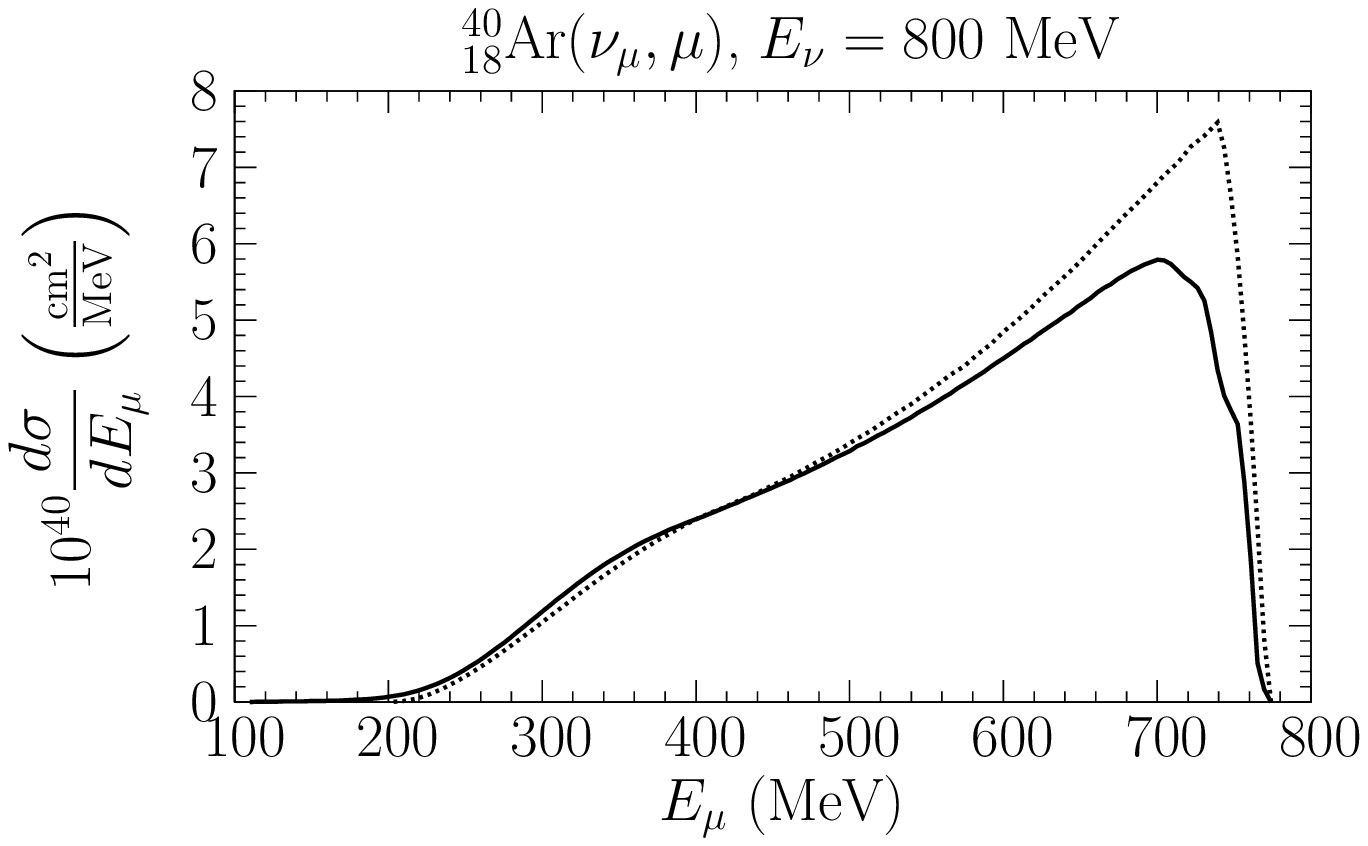}%
    \end{minipage}
\caption{\label{fig:Ar}Differential cross sections of argon. Comparison of our approach (solid line) to the FG model (dotted line). Left panel: The process $^{40}_{18}$Ar$(e,e')$. Experimental points are from~\cite{ref:Anghinolfi_Ar}. The value of momentum transfer at the peak is 371~MeV. Right panel: The process $^{40}_{18}$Ar$(\nu_\mu,\mu^-)$.}
\end{figure}

\paragraph{\bf Logic of our work}
The basic steps of our research were:
\begin{itemize}
\item{{Identification of the region in the energy transfer and momentum transfer ($\omega$, \n q) plane, from which most of the neutrino QE cross section comes from.}
    }
\item{{Selection of the electron scattering data with similar $\omega$ and~$\n q$.}
    }
\item{{Comparison of the results for oxygen obtained within our model to the available electron scattering data~\cite{ref:Anghinolfi,ref:O'Connell}} and the Benhar SF~\cite{ref:Benhar&Fabrocini&Fantoni&Sick,ref:Benhar&Farina&Nakamura}.
    }
\item{{Verification of accuracy for $^{40}_{20}$Ca$(e,e')$ and comparison with the results of Ref.~\cite{ref:Butkevich&Mikheyev}.}
    }
\end{itemize}
These two tests showed satisfactory agreement of our approach with the data, see Figs.~\ref{fig:Oxy} and \ref{fig:Ca}. Therefore the next step was taken: Application of the model to argon. The left panel of Fig.~\ref{fig:Ar} presents comparison with the few known experimental points for $^{40}_{18}$Ar$(e,e')$ scattering~\cite{ref:Anghinolfi_Ar}. Since the accuracy is comparable to the case of oxygen~\cite{ref:Ankowski&Sobczyk_GSF}, we proceed to our final goal and apply the new SF to neutrino scattering. The typical plot is presented in the right panel of Fig.~\ref{fig:Ar}.

\paragraph{\bf Discussion}
Only the QE scattering was considered. Thus for higher values of~$\omega$, some of the cross section is lacking. The oxygen nucleus provides the opportunity to compare our approximation with the more systematic Benhar SF, see Fig.~\ref{fig:Oxy}. The simplifying assumptions are responsible for slightly different shape of the QE peak obtained within our model, but the discrepancy is of the size of the error bars, and the achieved accuracy is good. Figure~\ref{fig:Ca} shows that for heavier calcium target, the agreement with data is better than for oxygen.

We conclude that the way we model SFs works fine in the region of $\omega$ and~$\n q$ relevant to $\sim 1$~GeV neutrino QE scattering. By adopting the standard extrapolation, we expect that the model tested for electrons describes neutrino scattering equally well. It allows us to presume that the accuracy of the neutrino cross section plot shown in the right panel of Fig.~\ref{fig:Ar} is similar to those presented for electron scattering.

In the case of all three targets, larger deviations from the electron scattering data occur for lower beam energies. This can be understood as an indication that when momentum transfer at the QE peak is lower than $\sim$400~MeV, the IA is no longer reliable.

\paragraph{\bf Acknowledgments}
The authors were supported by MNiSW under Grants No. 3735/H03/2006/31 (JTS,
AMA) and No. 3951/B/H03/2007/33 (AMA).


\bibliographystyle{aipproc}   

\end{document}